# Manifestation of kinetic-inductance in spectrally-narrow terahertz plasmon resonances in thin-film $Cd_3As_2$

Ashish Chanana[1,*], Neda Loftizadeh[2], Hugo O. Condori Quispe[1], Prashanth Gopalan[1], Joshua R. Winger[3], Steve Blair[1], Ajay Nahata[1], Vikram Deshpande[2], Michael A. Scarpulla[1,3], Berardi Sensale-Rodriguez[1,*]

[1] Department of Electrical and Computer Engineering, The University of Utah, Salt Lake City, UT 84112

[2] Department of Physics and Astronomy, The University of Utah, Salt Lake City, UT 84112

[3] Department of Materials Science and Engineering, The University of Utah, Salt Lake City, UT 84112

* E-mail: ashish.chanana@utah.edu ; berardi.sensale@utah.edu

## Abstract

Three-dimensional (3D) semimetals have been predicted and demonstrated to have a wide variety of interesting properties associated with its linear energy dispersion. In analogy to two-dimensional (2D) Dirac semimetals, such as graphene, $Cd_3As_2$, a 3D semimetal, has shown ultra-high mobility, large Fermi velocity, and has been hypothesized to support plasmons at terahertz frequencies. In this work, we demonstrate synthesis of high-quality large-area $Cd_3As_2$ thin-films through thermal evaporation as well as the experimental realization of plasmonic structures consisting of periodic arrays of $Cd_3As_2$ stripes. These arrays exhibit sharp resonances at terahertz frequencies with associated quality-factors (Q) as high as ~ 3.7. Such spectrally-narrow resonances can be understood on the basis of a large kinetic-inductance, resulting from a long momentum scattering time, which in our films can approach ~1 ps at room-temperature. Moreover, we demonstrate an ultrafast tunable response through excitation of photo-induced carriers in optical pump / terahertz probe experiments. Our results evidence that the intrinsic 3D nature of $Cd_3As_2$ provides for a very robust platform for terahertz plasmonic applications. Overall, our observations pave a way for the development of myriad terahertz (opto) electronic devices based on $Cd_3As_2$ and other 3D Dirac semimetals, benefiting from strong coupling of terahertz radiation, ultrafast transient response, magneto-plasmon properties, etc. Moreover, the long momentum scattering time, thus large kinetic inductance in $Cd_3As_2$, also holds enormous potential for the re-design of passive elements such as inductors and hence can have a profound impact in the field of RF integrated circuits.

Dirac semimetals, based on a peculiar quantum material with linear band dispersion and gapless electronic excitations, have been the subject of extensive fundamental research and are envisioned to facilitate profound technological advancements [1-5]. Dirac cone physics, in these materials, results in relativistic charge transport, suppression of backscattering, saturated light absorption, and electronic chirality; thus promises exciting advancements in applications such as low-power electronics, optoelectronics, quantum computing, and beyond [6,12]. Recent developments in graphene and topological insulators (TIs) have drawn attention towards the strong coupling of electromagnetic radiation as plasmons in these materials. This opens a new frontier of sub-wavelength confinement and strong light-matter interaction, enabling electron-plasma-wave devices and inducing non-linear effects [13-17]. These plasmons, in general, exhibit two distinct features: (*i*) in contrast to parabolic band materials, the plasma oscillations here are predominantly quantum mechanical in nature (since $\hbar$ appears as a leading term in the plasma dispersion relation). Furthermore, these charge oscillations cease to exist as the charge density (*n*) vanishes when the Fermi level is at the Dirac point ($\varepsilon_F = 0$). In addition, when compared to parabolic band materials, (*ii*) the linear energy dispersion in Dirac semimetals results in a lower-order dependence of the plasmon resonance frequency ($\omega_p$) on *n*, such that plasmon resonances predominantly lie at terahertz frequencies [18, 19]. The presence of strong coupling of terahertz radiation, in conjunction with facile tunable response and a long momentum scattering time ($\tau$) in these materials, has been envisioned to play a critical role in the burgeoning field of terahertz electronics and optoelectronics. In this regard, research efforts to-date have been mainly driven by studies of plasmons in two-dimensional (2D) graphene [20, 21]. Interestingly, long anticipated, three-dimensional (3D) analogues of graphene, i.e. 3D Dirac semimetals (3D-DSMs), have been recently experimentally demonstrated. Cadmium arsenide, $Cd_3As_2$, is a prototype of this class of materials [22-24]. Despite of their recent discovery, a variety of synthesis techniques have already been reported, and signatures of superior carrier transport, magneto transport and photo-carrier dynamics have been already demonstrated in these materials [25-29]. This naturally motivates further investigations on electromagnetic phenomena and light matter-interaction, such as Dirac plasmons. In this regard, the three-dimensional nature of 3D-DSMs results in a few distinctions from their more prevalently studied two-dimensional counterparts. Recent theoretical studies by Hofmann and Das Sarma [30] as well as Kharzeev *et al* [31] on plasmons in 3D-DSMs point out a fundamental difference in terms of their dispersion relations compared to those in 2D-DSMs. In 3D-DSMs, the plasma frequency is characterized by $\omega_p \propto \varepsilon_F \propto n^{1/3}$, while in 2D-DSMs, $\omega_p$ has a $n^{1/4}$ dependence [18-20]. In general, a particular charge density dependence of long wavelength plasmons in massless Dirac systems is of the form $\omega_p \propto n^{(D-1)/2D}$, where *D* is the dimensionality of the system [32]. Angle-resolved photoemission spectroscopy (ARPES) studies on $Cd_3As_2$ have revealed a Fermi velocity $v_F \cong 2\times10^6$ m/s which is approximately twice that of graphene [23]. Interestingly, this has a direct implication on plasmon characteristics; in general, higher $v_F$ (and longer $\tau$) produce plasmon resonances with larger quality-factor (Q) [18]. This situation is similar to what occurs in parabolic band materials, where low effective masses and long momentum scattering times are associated with high Q-factors in plasmonic resonances [33]. Another direct outcome of a long $\tau$ in DSMs is a large kinetic inductance ($L_k$) arising from the higher kinetic energy of carriers and given by $L_k = \tau/\sigma_0$ with $\sigma_0$ being the low-frequency limit of its AC conductivity. As can be understood from the previous equation, in conventional metals, with $\tau \sim$ fs and $\sigma_0 > 10^7$ S/m, kinetic inductance is typically negligible. Kinetic inductance in DSMs has been explored recently in graphene where spiral inductors fabricated using intercalated graphene sheets showed that the overall inductor footprint can be reduced by >1.5 times with respect to its metallic counterparts [34]. From this perspective, practical applications of kinetic inductance hold remarkable prospects for downscaling the size of radio-frequency and microwave circuits, where the large size of metal-based inductors has been a bottleneck. However, issues with large-area synthesis of high quality materials for such applications still looms, where the 2D nature of graphene has innate fabrication challenges [35, 36]. Benefitting from their bulk nature, 3D-DSMs are less susceptible to substrate and bulk phonon and scattering effects, degradation due to lithographic processes, etc. From a practical perspective, this can be beneficial for kinetic inductance -based devices, such as compact inductors. In addition, as previously mentioned, much like 2D-DSMs, plasmon resonances in 3D-DSMs inherently lie in the terahertz frequency range. Hence, the use of 3D-DSMs as a materials-platform for future microwave and terahertz devices could lead to many theorized applications, including electron plasma-wave based devices such as high responsivity resonant detectors [37], plasmonic terahertz sources [38, 39], etc. Another universal fingerprint of linear energy dispersion materials is the ultrafast recombination dynamics of photo-excited Dirac fermions, which could be exploited towards the design of ultrafast optoelectronic devices.

In this work, we present a direct demonstration of spectrally-narrow terahertz plasmon resonances in patterned $Cd_3As_2$ resonant plasmonic structures. This observation is attributed to the exceptionally long momentum scattering time measured in our films, and is a direct outcome of the large $v_F$ defining the Dirac dispersion of this material. We present a comprehensive study, starting with synthesis of polycrystalline thin films using thermal evaporation and detailing their structural, electrical, and terahertz characterization. Furthermore, we find that annealing of the as-deposited films under an inert environment results in reorientation and recrystallization of the films yielding larger grains and corresponding improvements of both DC and high-frequency electron transport properties. In order to effectively couple the incident terahertz radiation into terahertz plasmons, we patterned the films using a polymer-based delamination technique [40], devoid of any lithography or milling processes, thus preserving the properties of the original films. High quality factors, as high as Q ~ 3.7 ± 0.2, are in agreement with the modelled response derived from extracted Drude parameters and a low refractive index dielectric environment as provided by the quartz substrate. Systematic full-wave analysis enables us to elucidate the physical origin of these resonances and to explain the differences between the observed response and that of dispersion-free metals. Using time-resolved techniques, we also demonstrate strong photo absorption in our films, resulting from doubly degenerate linear bands in $Cd_3As_2$. These photo-excited carriers recombine in < 40 ps in accordance with previous reports in the literature [28, 29]. Excitation of photo-carriers in patterned samples leads to a dynamically tunable ultrafast plasmon response, which can enable future development myriad terahertz optoelectronic devices. Overall, our study introduces and highlights 3D-DSMs as prospective constituents for terahertz applications and beyond

For the purpose of our study, large area, continuous films of $Cd_3As_2$ were thermally evaporated from a lumped $Cd_3As_2$ source [American Elements, Product Code # CD-AS-05-L] on to quartz substrates under optimized conditions (as detailed in the *Methods* section). The major parameter affecting the grain size and structural properties was found to be the substrate temperature, where optimal chemical and structural properties were obtained at a substrate temperature of ~ 98° C (see *Supporting Information*, **Fig. S1**). **Figure 1** shows structural and compositional characterization of the polycrystalline films. As grown polycrystalline films, when annealed in Ar gas at 450 °C, recrystallized to form larger grains with typical lateral dimensions of about 400 nm (see **Fig. 1 (a)**). In annealed samples, we observed sharp crystalline grain facets. Consequently, X-ray diffraction (XRD) analysis [Bruker D2 PHASER] revealed a family of diffraction peaks corresponding to the tetrahedral structure, *I4₁cd*, of $Cd_3As_2$ (see **Fig. 1(b)**). This XRD data is in good agreement with powder XRD data, published elsewhere [41]. We confirmed molecular composition using Raman spectroscopy, using a 488 nm excitation, where as depicted in **Fig. 1(c-d)** phonon vibrational modes at ~ 194 cm$^{-1}$ and ~ 249 cm$^{-1}$ were observed [42, 43]. To validate the oxidation states of elements, we measured binding energy in X-ray photoemission (XPS) spectrum from Cd 3d and As 3d peaks, favoring oxidation state corresponding to $Cd_3As_2$. We also observed partial oxidation of films, as supported by XPS data (see *Supporting Information*, **Fig. S2**), where oxidation of the top ~ 5-10 nm of the film was observed. However, oxidation of the film was found to be self-limiting.

Linear dispersion in energy-momentum (*E-k*) space manifests as a unique dependence of plasmon frequencies on the number of carriers: as opposed to the characteristic square-root dependence ($\omega_p \propto n^{1/2}$) for parabolic band semiconductors, DSMs exhibit lower-order dependences [30, 31]. The schematic representation in **Fig. 2(a)** captures the effect of intra-band transitions excited by THz probe fields producing a free carrier (Drude) response at terahertz frequencies. The electromagnetic response of thin-films (optical thickness << λ) can be understood through an effective surface conductivity approach [44]. Here, assuming a Drude model, i.e. $\sigma(\omega) = \sigma_0/(1 + i\omega\tau)$, the frequency dependent complex admittance, $Y = 1/\sigma(\omega)$, representing the film is modelled as the series connection of a series resistance $R = 1/\sigma_0$ and the kinetic inductance $L_k = \tau/\sigma_0$ as schematically depicted in **Fig. 2(b)**. Terahertz transmission through our thin-films was measured using THz time domain spectroscopy (THz-TDS). We extracted the complex conductivity using Fresnel coefficients from the transmission through the films [44, 45]. By fitting the experimentally extracted frequency-dependent conductivity depicted in **Fig. 2(c)** to a Drude model and employing a nominal thickness for the film of ~220 nm, as confirmed by SEM, we extracted the following parameters: $\sigma_0 = 3.5 \pm 0.5 \times 10^5$ S/m and $8 \pm 0.3 \times 10^5$ S/m before and after annealing, respectively. The corresponding Drude scattering times, were found to be $\tau \sim 0.35 \pm 0.05$ ps and $\sim 0.70 \pm 0.05$ ps, respectively. In general, across several samples, synthesized over time, our process yielded $\sigma_0$ ranging between $\sim 1$ to $\sim 5\times 10^5$ S/m (before annealing) and $\sim 3$ to $\sim 10\times 10^5$ S/m (after annealing); with relaxation times ranging between $\sim 0.3$ and $\sim 0.5$ ps (before annealing) and $\sim 0.7$ to $\sim 1.0$ ps (after annealing). For illustrative purposes, depicted in the inset of **Fig. 2(c)** is the extracted conductivity for a particular film with extracted relaxation time $\sim 1$ ps. The deviations in the extracted parameters mainly stem from variations in materials quality across growths, whereas the uncertainty arises from spatial variations in film thickness as well as on the limited bandwidth of terahertz spectroscopy system. In addition, we extracted DC conductivity and mobility from Hall effect measurements (see **Fig. 2(d)**). Measurements were performed on structures using bottom gold contacts, which bypass the problem of oxidation of the top film surface. Here, assuming a nominal thickness of $\sim 220$ nm, DC conductivity and mobility for the un-annealed film were found to be $\sim 5.5\times 10^4$ S/m and $\sim 400$ cm$^2$/V.s, respectively; while annealing of the film increased these values to $\sim 3.4\times 10^5$ S/m and $\sim 1,000$ cm$^2$/V.s, respectively. This observed increase in DC-extracted conductivity and mobility after annealing are consistent with the corresponding increases observed in THz-extracted parameters. However, the THz-extracted zero-frequency conductivity levels are larger than the DC-extracted ones. The differences in conductivity levels obtained from terahertz and DC measurements result from the distinct probe lengths at which transport is measured in these measurements. The transverse characteristic length at which terahertz illumination probes electronic transport is on the nm scale [45-47]. From this perspective, in the case of terahertz measurements, charge transport at high frequencies, is largely, on average, limited within grains, thus, micrometer scale effects such as grain boundaries do not affect transport. However, in the case of DC measurements, transport is measured using longitudinal and transverse Hall bars, placed 20 µm apart, in the Van der Pauw geometry; hence conductivity is measured across several grains. From Hall measurements, the carrier density was estimated to be $1.6 \times 10^{19}$ cm$^{-3}$ and $9.8 \times 10^{18}$ cm$^{-3}$ in annealed and un-annealed films, respectively. Such high doping densities are common for films deposited using thermal evaporation of Cd$_3$As$_2$, while more moderate n-type doping is common in films prepared using other techniques [49, 50]. As per literature reports, based on DFT modelling, similar doping concentrations were estimated to form a Fermi level of ~ 200 meV, where, carriers occupy the linear regime [28] and thus showcase Dirac phenomena.

Next, we proceed to characterize localized plasmon resonances in thin-films of Cd$_3$As$_2$. This required patterning the films to enable coupling of incident terahertz radiation into plasmon modes. This has been quite a challenge with 2D sheets of graphene where extensive efforts have been made to reduce the degradation of mobility and $\tau$ due to lithographic process [36]. Here, we circumvent the need for developing an etching process for Cd$_3$As$_2$ by adopting our unique approach of depositing on to patterned polymer films and subsequently delaminating the polymer, as demonstrated recently [40]. The non-destructive (solvent and wet-process free) approach and the fact that Cd$_3$As$_2$ deposition is reserved as the last step before polymer peel-off, ensures the same quality of Cd$_3$As$_2$ as obtained in un-patterned films. The fabrication process is illustrated schematically in **Fig. 3**. We patterned Cd$_3$As$_2$ into ribbons as these are resonant structures for which quantitative models are widely available [51-53]. The patterns were designed to exhibit resonances below 1.5 THz. The scalable response of plasma frequency on the patterned structures could be approximated, under quasi-static limit, by solving Maxwell's equation for polarization fields along the boundaries. Under a thin film approximation, provided that the thickness of the film is much smaller than a skin depth, for stripes of width *W*, the expression for the first plasmonic resonance ($\omega_p$) is given by [51-53]:

$$\omega_p \propto \sqrt{\frac{\pi \sigma_0}{\epsilon_{avg} \tau W}}, \tag{1}$$

where $\varepsilon_{avg}$ is the average permittivity surrounding the stripes. Eqn. (1) is formally valid for a "zero-thickness" conductive sheet of zero-frequency dynamic sheet conductivity $\sigma_0$, i.e. $\sigma_0$ has units of S. It is to be noted that such is not the "exact" case in our samples. However, the general trends with respect to geometric dimensions and momentum scattering time predicted by Eqn. (1) still hold in our experimental situation. Similar square-root dependence of resonance frequency on the structure dimensions and momentum scattering time has been also established for disk geometries [54].

In order to account for the effect of film-thickness, full-wave numerical simulations were performed to understand the effects of $\tau$, $\sigma_0$, and *W* in the terahertz transmission through periodic stripe patterns (with conductivity modelled through a Drude model). To provide for a systematic analysis, one parameter was altered at each time while the others remained constant. First, by assuming $\tau = 0$ and *W* = 100 µm, we analyzed the effect of varying $\sigma_0$. This situation corresponds to the film behaving as a "dispersion-free lossy conductor"; simulations were performed for $\sigma_0$ = $10^3$, $10^5$, and $10^7$ S/m. These levels represent conservative estimates for conductivity in semiconductors, semimetals, and metals, respectively. As depicted in **Fig. 4(a)**, for a terahertz excitation polarized perpendicular to the stripes, three important features are observed: (*i*) at low frequencies, transmission drops following the characteristic low pass filter response of a capacitive grid (*RC* roll-off); (*ii*) this drop is then followed by a subsequent sharp increase in transmission, where transmission peaks at a geometrically defined extraordinary optical transmission (EOT) resonance, $\omega_0$; (*iii*) when altering the conductivity of the film the position of the resonance, $\omega_0$, which is geometrically defined, remains constant. Next, by assuming $\sigma_0 = 10^5$ S/m and $W = 100$ μm, we analyzed the effect of varying $\tau$. **Figure 4(b)** depicts color-maps of the transmission spectra. As $\tau$ is increased, the following trends are observed: (*i*) a transition of the spectrum from geometrically defined resonances into several branches of localized plasmonic modes, which are manifested by spectrally-narrow dips in transmission; (*ii*) a spectral-narrowing of the resonances (larger Q-factor); as well as (*iii*) a red-shift in their position. It is interesting to note that the observed red-shift and narrowing of the resonances can be explained on basis of a series *RLC* equivalent circuit, where the resonance frequency follows $\omega_p \propto \frac{1}{\sqrt{C(L_g+L_k)}}$ and the quality factor follows $Q \propto \sqrt{\frac{L_g+L_k}{C}}$. From this point of view, as $\tau$ is increased, the kinetic-inductance ($L_k$) can become comparable or larger than the geometric inductance ($L_g$). Therefore, the resulting resonances become dominated by $L_k$ and thus constitute a direct manifestation of electron inertia. It is to be noted that dips in transmission would also appear in dispersion-free metal gratings, i.e. $\tau = 0$ ps, however these manifest as broad valleys formed between EOT peaks imposed with base-level terahertz transmission determined by the film conductivity. In contrast, the spectrally narrow resonance dips observed in this study, which are associated with plasmon coupling, are only possible in case of conductivity dispersion imparted by a long momentum scattering time. Another interesting observation in **Fig. 4(b)** is that when increasing $\tau$, we observe two branches appearing near the second order resonance branch, $\omega_{p,II}$ : a pure plasmonic resonance, $\omega_{p,IIa}$, and an hybrid mode, $\omega_{p,IIb}$; with both branches associated with spectrally-narrow resonances. However, whereas the extinction at $\omega_{p,IIa}$ slowly varies when altering $\tau$, the extinction at $\omega_{p,II}$ sharply decreases when $\tau$ is increased and $\omega_{p,II}$ approaches $\omega_0$. Finally, by assuming $\sigma_0 = 10^5$ S/m as well as $\tau = 0$ (**Fig. 4(c)**) and $\tau = 0.5$ ps (**Fig. 4(d)**), we analyzed the effect of varying $W$. For a typical case representing the 3D-DSMs analyzed in this study (corresponding to $\tau = 0.5$ ps), a $W^{-0.5}$ dependence is observed (see inset in **Fig. 4(d)**). The latter observation holds for both the $\omega_{p,I}$ and $\omega_{p,II}$ branches. Our analysis shows that the plasmonic modes in 3D-DSMs are not only characterized by spectrally narrow dips in transmission (in contrast to broad *LC* resonances as is the case in dispersion-free metals), but also by: (a) a dynamic behavior (with regard to the resonance position) when altering either $\tau$ or $\sigma$, as well as by (b) a characteristic $\omega_p \propto W^{-0.5}$ dependence. The strong dependence of the resonances on $\tau$ is a manifestation of kinetic-inductance and can enable either: (*i*) resonances at much smaller frequencies ($\omega_{p,I}$) than the geometrically defined LC resonances in the metal case, or (*ii*) resonances with very large Q-factors ($\omega_{p,II}$) near $\omega_0$. Depicted in **Fig. 5** is the measured transmission spectra through a stripe-patterned film with $W = 100$ μm. A good qualitative agreement is observed between our measured data and full-wave electromagnetic simulations in the sense that multiple plasmonic resonances (dip in transmission) are observed in-between EOT peaks. Two EOT and four plasmonic resonances are identified in the measured spectra. In general, higher order plasmonic modes, i.e. higher frequency resonances, are observed to have larger quality factors.

After having characterized the general features of the resonances, we now measured transmission through a set of samples with varying properties. The first effect that we explore is the role of polarization on the transmission through a stripe-patterned film. For this purpose, we look at the response at the two orthogonal polarizations as depicted in **Fig. 6(a)**. When the incident THz beam is polarized parallel to the stripes, a monotonically increasing transmission, consistent with a Drude response, is observed. However, when the incident polarization is perpendicular to the stipes, the response shows a resonant (Lorentzian) response, the origin of which has been discussed in the previous section. From the non-resonant transmission, a relaxation time of ~0.5 ps is extracted, which is in close agreement with the results observed from transmission measurements through un-patterned films. The effect of polarization can be simply understood in-basis of equivalent circuit models. When the terahertz field is polarized parallel to the direction of the stripes, i.e. $E_{THz} = E_{\parallel}$, the response of the structure essentially follows a *RL* circuit and therefore transmission increases with frequency at a rate set by $L_k$ and thus directly dependent on $\tau$. When the terahertz field is polarized perpendicular to the direction of the stripes, i.e. $E_{THz} = E_{\perp}$, the response could be represented by a resonant series *RLC* circuit as discussed in previous sections. We model the resonant response of measured resonances as a damped-Lorentzian oscillator, where the measured transmission response was fitted to:

$$\frac{A}{(\omega-\omega_p)^2 - (\Gamma_{res}/2)^2}. \qquad (2)$$

Here $\Gamma_{res}$ represents the linewidth of the resonance, and A its magnitude. Lorentzian fits of our data to Eqn. (2) are plotted as dashed curves in **Fig. 6(a)**. We extracted the Q-factor of the resonance to be ~ 2.45 ± 0.1 at ~0.75 THz. This Q is substantially larger than what has been previously reported in graphene [13, 14] and is on the order of Q ~ $\omega_p \tau$. This large Q might also indicate an effectively longer $\tau$ for longitudinal, i.e. plasmonic, electromagnetic wave propagation than what is the case under normal incidence transmission. Discrepancies with regard to constitutive parameters affecting normal incidence and longitudinal measurements have been previously reported in metals [55]. Furthermore, a broadening of the EOT peaks could also in part be contributing to a further experimental enhancement in Q-factor of adjacent resonant dips. We now explain the effect of annealing the samples on their terahertz response. When comparing samples before and after annealing, we observe an increase in Q, an increase in extinction, as well as a blue-shift of $\omega_p$ (as depicted in **Fig. 6(b)**), which are consistent with a longer $\tau$, a larger $\sigma_0$, and a larger *n*, respectively. It is to be noted that the charge density, *n*, governs the position of the resonance, while τ defines its strength. The observed Q in this case is even larger, ~3.7 ± 0.2, which is consistent with an increase of ~2X in τ after annealing as observed through normal incidence measurements through bulk films (see **Fig. 2(b)**). To the best of our knowledge this Q-factor is among the largest reported to date in plasmonic structures at this frequency range. Finally, we also explored the effect of altering the geometric dimensions in our samples. Depicted in **Fig. 6(c)** is the effect of altering *W*. The two curves presented correspond to transmission through stripe-patterned samples having *W* = 100 μm, *L* = 200 μm and *W* = 70 μm, *L* = 140 μm, respectively. It is observed that decreasing the geometric dimensions leads to a blue shift of the resonance, which is consistent with the qualitative trend predicted by Eqn. (1); however due to a large variability across samples resulting from our deposition process, it is not possible for us to conclude with a particular scaling law. Overall, our observations establish the first direct demonstration of terahertz plasmons in a 3D-DSM, namely $Cd_3As_2$.

In order to employ 3D-DSMs for terahertz plasmonics, active modulation of carrier density could play a crucial role. This is often achieved by electrostatic gating of 2D-DSMs. We explore the possibility of modulating the carrier density by injecting photo-carriers. In this regard, gapless and linear dispersion materials have also been associated with unique photo physics, marking an ultrafast recombination of carriers [56]. Hence, here, we employ time-resolved optical pump measurements to demonstrate an ultrafast carrier modulation in $Cd_3As_2$. The measured response was marked by strong optical absorption and carrier relaxation in < 40 ps, with optical excitation at 800 nm. **Figure 7(a)** illustrates the allowed optical transitions in a typical gapless material where only transitions arising from excitations with energy ≥ $\varepsilon_F$ are allowed. Hence, assuming the $\varepsilon_F$ ~ 200 meV in annealed films, 800 nm excitation would excite carriers to energy of ~750 meV above the Dirac point. The measured transmission through un-patterned films shows a strong modulation (**Fig. 7(b)**). Using similar pump excitations on stripe-patterned samples demonstrates a noticeable tuning of the plasmon resonance (**Fig. 7(c)**), where enhanced extinction and blue-shift of the resonance displays the expected trend with carrier injection (as discussed earlier in the manuscript). Thus, we also established optical pump excitation as an effective route to ultrafast tunable terahertz plasmon response in $Cd_3As_2$. However, given the high doping concentration of the as–grown films, the induced change upon photoexcitation is limited. As discussed earlier, the high carrier density has been an inherent problem associated with the material, where arsenic vacancies impart an unintentional n-type doping to the material. This limitation could be overcome, in the future, through optimized growth conditions [48, 49].

The exploration and practical application of terahertz plasmons in 2D-DSMs has become an attractive scientific topic following the discovery of graphene. Recently, Dirac plasmons in topological insulators were also demonstrated [15]. Furthermore, terahertz plasmonics in non-Dirac materials including superconductors and high-mobility semiconductors have also been widely demonstrated (e.g. [54, 57-59]). Our work introduces another potential candidate, $Cd_3As_2$, to the "library" of materials for terahertz plasmonics. The first observation of Dirac plasmons in graphene was reported by Ju *et al.* [13], which was followed up by many other groups e.g. [14, 16, 60-62]. Recently, Daniels *et al.* reported narrow plasmons resonances in epitaxial grown graphene on SiC (n ~ 3.2) where a high Q-factor of ~ 1.2 was observed at a resonance frequency of 1.7 THz [53]. Having a similar momentum relaxation time, our samples can display higher quality factors, even in an ultra-thin film form, as we employed a substrate with a lower dielectric constant (n ~ 2.1) [61]. Furthermore, in bulk form, our examples exhibit multiple spectrally-narrow resonances originating from high-order as well as hybrid modes. At this end, in **Fig. 8**, we benchmark our results (using Q-factor as the criterion) against published values of terahertz plasmons in Dirac and non-Dirac materials [13, 14, 16, 53, 57-63 ]. **Figure 8** summarizes the Q-factors associated with plasmon resonances observed in Dirac and parabolic band materials, based on published studies. In general, plasmon resonances in graphene [14, 53] and two dimensional electron gases in high mobility semiconductors [54, 58] have been reported to have the highest Q-factors. In comparison, here we demonstrate strong resonances in a 3D-DSM where geometrically defined resonances transform into spectrally-narrow resonances as a result of the contribution of a large kinetic inductance.

In conclusion, we reported a strong terahertz plasmon response in $Cd_3As_2$, a 3D-DSM. The long momentum scattering time in thermally evaporated polycrystalline films enables spectrally narrow resonances. Furthermore, an ultrafast tunable response is demonstrated through excitation of photo-induced carriers. Our observations can pave a way for the development of myriad terahertz optoelectronic devices based on $Cd_3As_2$, benefiting from strong coupling of terahertz radiation, ultrafast transient response, magneto-plasmon properties, etc. Moreover, large kinetic inductance associated with long momentum scattering time, in $Cd_3As_2$ also holds enormous potential in the field of RF integrated circuits.

**Methods**

Sample Preparation: Thin film $Cd_3As_2$ were thermally evaporated from commercial $Cd_3As_2$ lumped source materials (American Elements, product # CD-AS-05-L), where crystalline growth was observed at an optimized substrate temperature. During the deposition the chamber pressure was maintained at $3 \times 10^{-5}$ Torr and the films were deposited at an average rate of 10.5 A/sec. We found the optimal substrate temperature to be in the range of 95-100C; lower temperatures did not yield conductive films while at higher temperatures the film did not adhere to the substrate. Annealing of the films was performed in a three-zone tube furnace under an inert atmosphere of Ar gas, where the temperature of the first zone (corresponding to the sample placement) was gradually (4° C/ min) heated to 450° C. The sample was maintained at this temperature for ~ 2 hours and was slowly cooled down with Ar flow to room temperature at about 2° C/ min.

Terahertz measurements: Ultrafast optical pulses were generated by a 1 KHz amplified Ti-sapphire laser (4 mJ, 85 fs pulses) system, where majority of the radiation is used for optical excitation of carriers while rest was employed for time domain terahertz spectroscopy, THz-TDS measurements. Linear spectroscopy was performed using a conventional THz-TDS setup using optical rectification and electro-optic sampling for generation and detection of terahertz radiation via 1 mm ZnTe [110] crystal. Time-resolved optical-pump terahertz-probe (OPTP) measurements were performed by varying the time delay between the optical pump and THz-probe pulses using a mechanical stage. The frequency dependent terahertz spectrum was obtained at different time delays after excitation were made by fixing the optical pump at a desired position and sampling the THz probe using a second delay stage.

**Acknowledgements**

The authors thank Dr. Paolo Perez for help with analyzing the XPS data. The authors acknowledge the support from the NSF MRSEC program at the University of Utah under Grant No. DMR 1121252 and from the NSF CAREER Award No. #1351389 (monitored by Dimitris Pavlidis). This work was performed in part at the Utah Nanofab sponsored by the College of Engineering, Office of the Vice President for Research, and the Utah Science Technology and Research (USTAR) initiative of the State of Utah.

**FIGURES:**

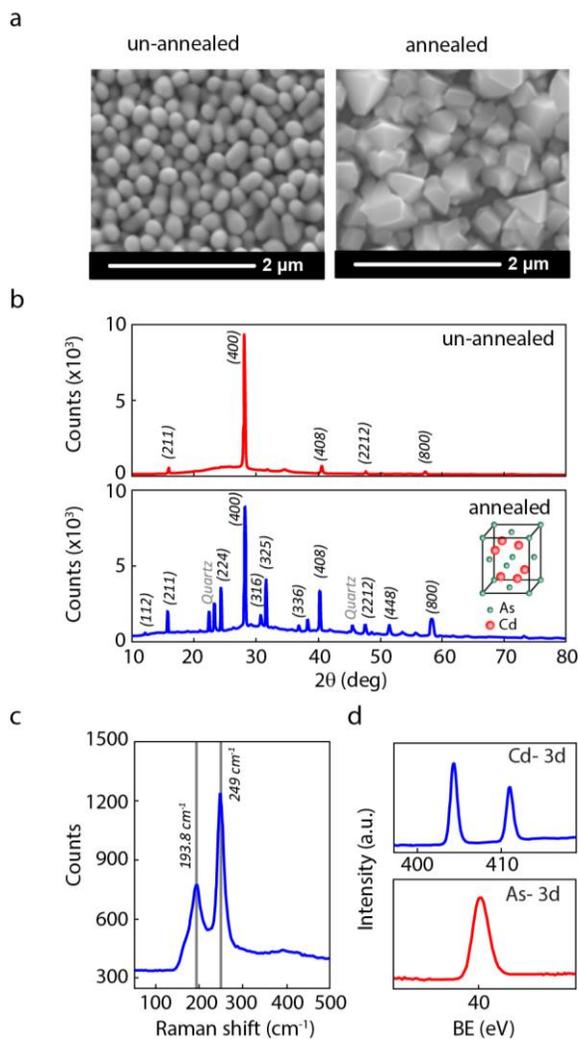

**Figure 1.** *Characterization of thin film Cd$_3$As$_2$ prepared using thermal evaporation.* **(a)** Representative SEM images of a un-annealed 400 nm thick Cd$_3$As$_2$ film (left) and annealed (right). In general, annealing results in sharper grain facets and larger grain domains. **(b)** Corresponding 2-theta XRD spectrum showing the diffraction peaks from polycrystalline film corresponding to tetrahedral lattice unit cell of Cd$_3$As$_2$, as shown in the inset. **(c)** and **(d)** depict Raman and XPS data obtained on the annealed films, confirming the molecular bonds and oxidation states of Cd and As in the desired form.

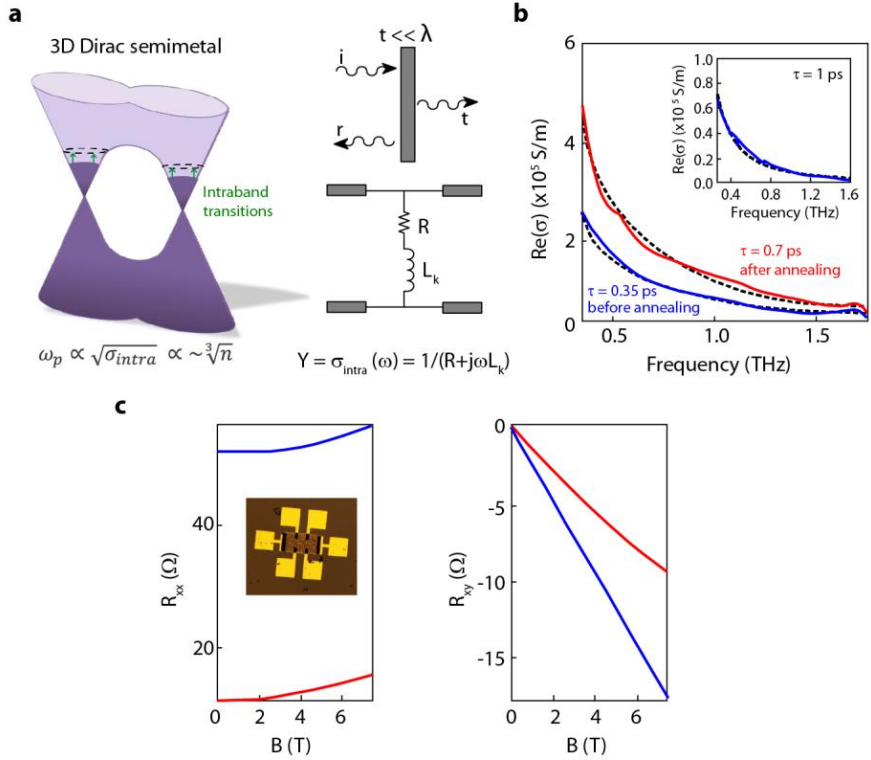

**Figure 2.** *Terahertz and DC conductivity in $Cd_3As_2$ films.* **(a)** *Right:* Schematic representation of intra-band transitions at low-energy terahertz frequencies. *Left:* transmission line model for transmission through thin film ($t \ll \lambda$), where intraband conductivity could be modelled as series resistance, R, and kinetic inductance, $L_k$, providing the Drude dispersion. **(b)** Extracted terahertz conductivity in the 0.4 – 1.6 THz frequency range for a film after (blue) and before (red) annealing. The dashed curves correspond to fitting of the extracted data to a Drude model. Across several analyzed samples, our process yields $\sigma_0$ ranging between ~1 to ~5×10$^5$ S/m (before annealing) and ~3 to ~10×10$^5$ S/m (after annealing); with relaxation times ranging between ~0.3 and ~0.5 ps (before annealing) and ~0.7 to ~1.0 ps (after annealing). **(c)** Hall measurements before (red) and after (blue) annealing. In general, low-frequency conductivity and Drude scattering time as well as DC conductivity were found to increase after annealing of the films.

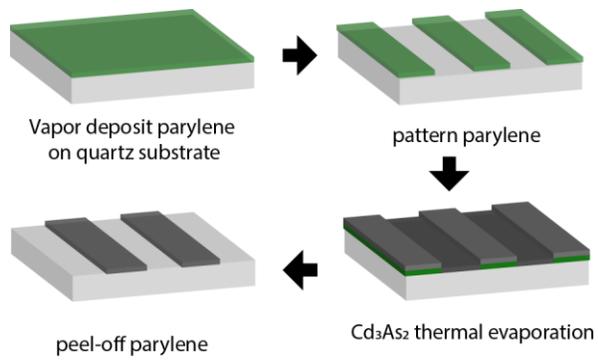

**Figure 3.** *Schematic illustration of our patterning process via polymer delamination technique.* Details of fabrication procedure is presented in Ref [39].

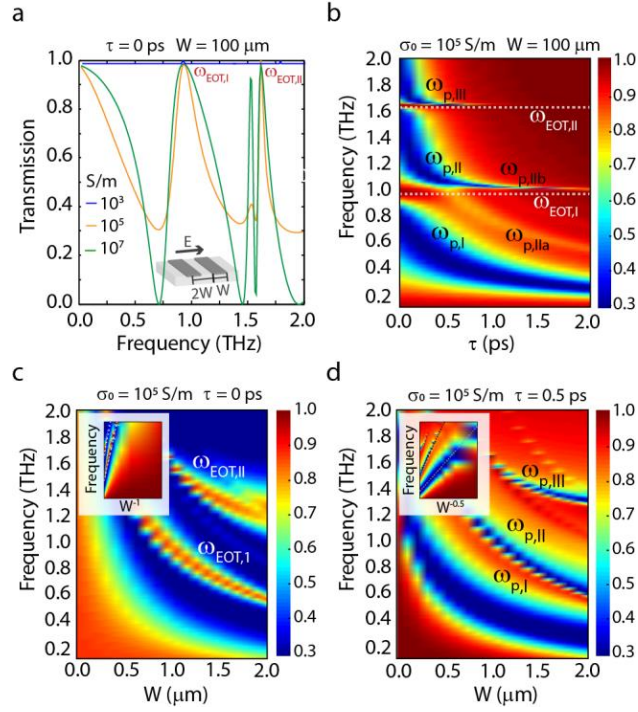

**Figure 4.** *Terahertz plasmonic response in patterned $Cd_3As_2$ films.* **(a)** Simulated transmission spectra for a dispersion-less metal lossy metallic film patterned into stripes; LC and EOT resonances are observed as minima and maxima in the transmission spectra. **(b)** Transmission versus frequency as a function of τ. A red-shift of the LC resonances, together with appearance of higher order and hybrid modes is observed for long τ. These effects are a manifestation of a large kinetic inductance, which in this case dictates the response of the film. **(c)** and **(d)** geometric scaling of the resonances for (c) τ = 0 and (d) τ = 0.5 ps. Characteristic $W^{-0.5}$ dependence on resonance frequency for plasmon resonances was observed.

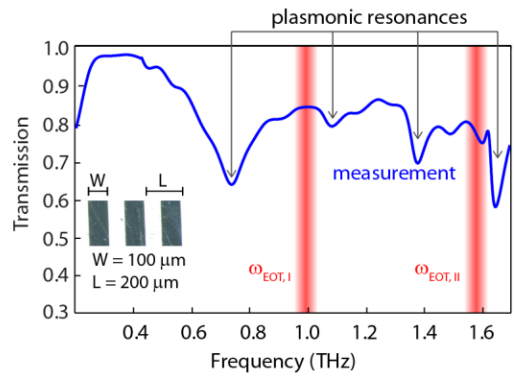

**Figure 5.** *Experimental characterization of plasmon resonances.* Transmission response of patterned stripe array showing multiple plasmon resonances (dips in transmission) and EOT peaks. The vertical shaded lines correspond to the simulated position of the EOT peaks.

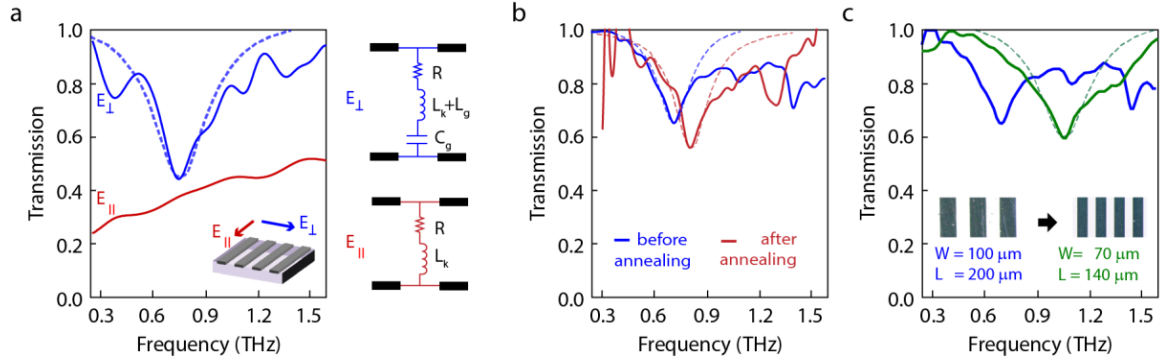

**Figure 6.** *Experimental analysis of geometric and material parameters.* **(a)** Effect of polarization. Transmission spectra through $Cd_3As_2$ stripes for incident terahertz beam polarization perpendicular (blue) and parallel (red) to the stripes. Whereas parallel polarization yields a Drude response, perpendicular polarization results in a Lorentzian response characteristic of a plasmonic resonance. The dashed lines indicate fits of the measured data to Drude and Lorentzian models. The schematic to the right depicts through an equivalent circuit representation the response along two orthogonal polarizations. Transmission response perpendicular to the stripes could be modelled with a series R-L-C circuit, where $C_g$ and $L_g$ are defined by geometry, while the total L contains contribution from geometric as well as kinetic inductance. **(b)** Effect of annealing. The blue and red curve represent transmission through a sample before and after annealing, respectively. Resonance blue shift, larger Q-factor, and larger extinction is observed after annealing; which is consistent with a longer τ as observed from transmission through un-patterned films and larger *n* as observed in Hall measurements. **(c)** Effect of decreasing the geometric dimensions. The blue and red curves correspond to transmission through samples having W = 100 μm, L = 220 μm and W = 70 μm, L = 140 μm, respectively. Overall, decreasing the geometric dimensions leads to a blue shift of the resonance.

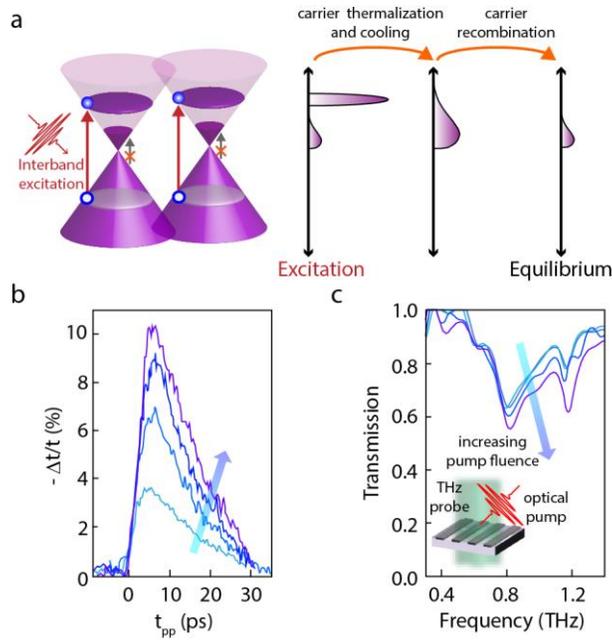

**Figure 7.** *Ultrafast carrier density modulation and plasmon tuning in Cd$_3$As$_2$.* **(a)** Schematic representation of optical pump excitation of carriers (left) and ultrafast carrier relaxation pathways (right) in Cd$_3$As$_2$. **(b)** Measured ultrafast carrier recombination in Cd$_3$As$_2$ films at different pump fluences; ultrafast response (< 30 ps) is observed, which makes this material attractive for ultrafast terahertz opto-electronic applications. **(c)** Plasmon resonance tuning in a stripe-patterned sample. The arrow marks the trend with increasing pump fluence, blue shift of the resonances as well as larger extinction is observed. The pump fluence levels in (b) and (c) are 35, 70, 90, and 150 μJ/cm$^2$.

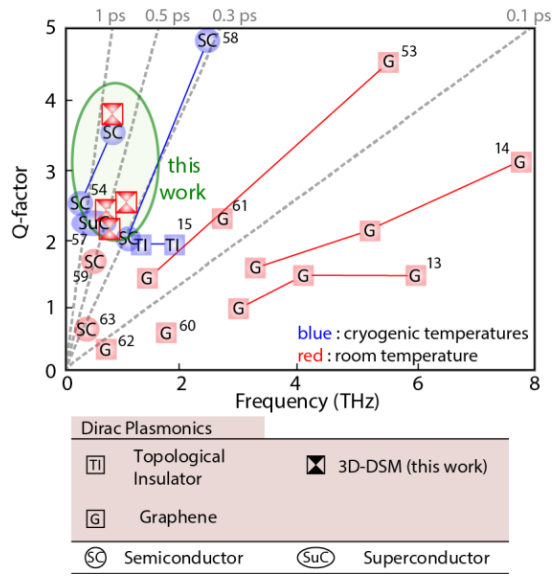

**Figure 8.** *Benchmarked Q-factors of terahertz plasmons with previous studies in the literature* (semiconductors, superconductors, topological insulators, and semi-metals). As a result of the long $\tau$ in our samples, the resonances observed in our work are among the narrowest in the 0 to ~1 THz spectral range. Dashed lines correspond to constant $\omega_p \times \tau$ product.